\documentclass[journal]{IEEEtran}
\IEEEoverridecommandlockouts
\usepackage{graphicx,amsmath,amssymb,xcolor,booktabs,url,array}
\usepackage[hidelinks]{hyperref}
\usepackage{cite}
\newcommand{\PL}{\mathrm{PL}}
\newcommand{\LoS}{\mathrm{LoS}}
\newcommand{\NLoS}{\mathrm{NLoS}}
\newcommand{\Deta}{\Delta_{\eta}}

\begin{document}

\title{Cost of Decibel-Domain Averaging of Line-of-Sight\\States in Aerial Coverage Analysis}

\author{İsmail G\"uven\c{c},~\IEEEmembership{Fellow,~IEEE}%
\thanks{This work was supported in part by the NSF award CNS-2450593.}%
\thanks{I. G\"uven\c{c} is with the Department of Electrical and Computer Engineering, North Carolina State University, Raleigh, NC 27606 USA (e-mail: iguvenc@ncsu.edu).}}

\maketitle

\begin{abstract}
Uncrewed aerial vehicle (UAV) coverage studies routinely average the line-of-sight (LoS) and non-line-of-sight (NLoS) path loss in the decibel domain, weighted by their probabilities. Since the conditional path-loss distribution is bimodal, the resulting mean path loss describes no physical link. We give the error of this mean path loss, relative to the path loss that corresponds to the mean received power, in closed form. We quantify it by ray tracing an ITU-R P.1410 city, supplying the averaged model with the exact LoS probability so that no parameter error contributes. Our numerical results show that the averaged path loss is 6.6~dB higher than the path loss of the mean received power and lies within 2~dB of only 6.2\% of the links. It moves the coverage-optimal altitude from 1764~m to 778~m and understates the outage probability by up to 0.198.
\end{abstract}


\section{Introduction}

\IEEEPARstart{T}{he} elevation-dependent line-of-sight (LoS) probability model is the standard tool for aerial coverage analysis, and it underlies most of the placement and deployment literature for aerial base stations~\cite{zeng2019}. For an elevation angle $\theta$ in degrees, this probability takes the sigmoid form
\begin{equation}
P_{\LoS}(\theta)=\frac{1}{1+a\exp\!\left(-b\left[\theta-a\right]\right)},
\label{eq:plos}
\end{equation}
where the environment-dependent constants $a$ and $b$ are fitted to a statistical city built from the ITU-R P.1410 built-up parameters~\cite{alhourani2014wcl,itu1410}, for example, $(a,b)=(9.61,0.16)$ for an urban city~\cite{alhourani2014wcl,boryaliniz2016}. The companion path-loss model adds an excess loss $\eta_{\LoS}$ or $\eta_{\NLoS}$ to the free-space path loss $\PL_{\mathrm{FS}}$ according to the link's state, line-of-sight or non-line-of-sight (NLoS)~\cite{alhourani2014wcl}. This construction is sound and widely validated, and this letter does not question it.

\begin{figure*}[!t]
\centering
\includegraphics[width=\textwidth]{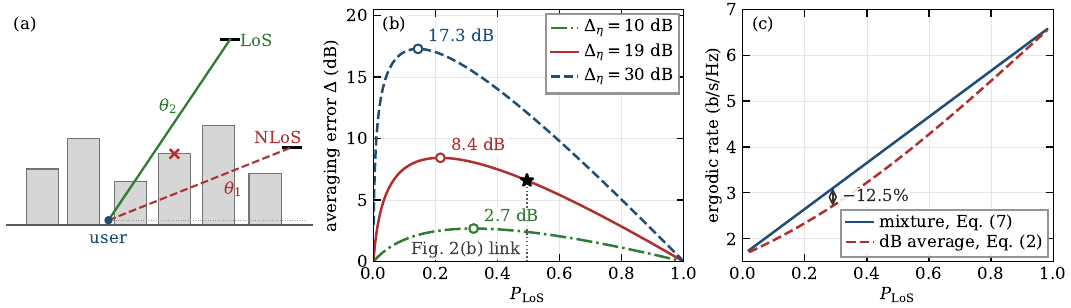}
\caption{Averaging error of a two-state link. (a) A ground user sees the aerial platform either over the rooftops or through them, so the link is either LoS or NLoS, and the elevation angle decides which. (b) Averaging error \eqref{eq:gap} against $P_{\LoS}$ for $\eta_{\LoS}=1$~dB and three mode separations $\Deta=\eta_{\NLoS}-\eta_{\LoS}$, with the maximum of each marked. The star and the dotted line mark the representative link of Fig.~\ref{fig:L3}(b), where $\Delta=6.6$~dB. (c) Ergodic spectral efficiency from the mixture \eqref{eq:rate} and from the averaged path loss \eqref{eq:dbavg}, at an LoS signal-to-noise ratio of 20~dB with $\sigma_{\LoS}=4$~dB and $\sigma_{\NLoS}=8$~dB. The arrow marks the largest relative error, $-12.5\%$ at $P_{\LoS}=0.29$.}
\label{fig:concept}
\vspace{-0.8\baselineskip}
\end{figure*}

\begin{figure*}[!t]
\centering
\includegraphics[width=\textwidth]{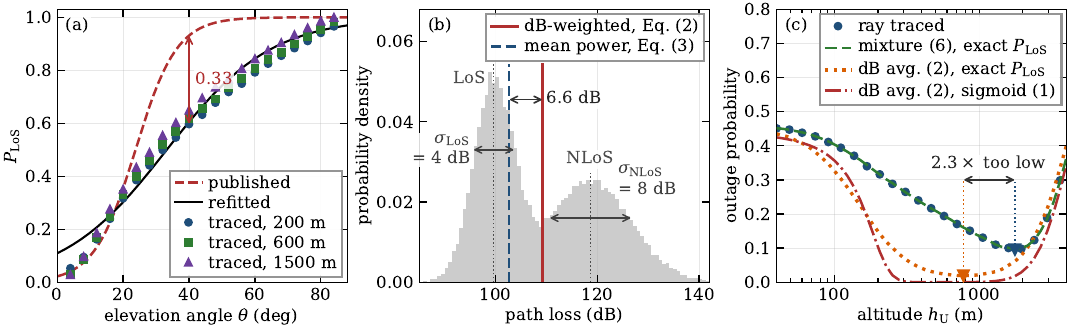}
\caption{Ray-traced evaluation in an ITU-R P.1410 urban city with $\alpha=0.3$, $\beta=500$~buildings/km$^{2}$, and $\gamma=15$~m. (a) LoS probability against elevation angle at three horizontal ranges, with the published urban sigmoid \cite{alhourani2014wcl,boryaliniz2016} and the same form refitted to this city; the arrow marks the largest error of the published pair. (b) Path-loss distribution for a link at 500~m range and 300~m altitude, with the decibel-weighted mean path loss \eqref{eq:dbavg} and the path loss of the mean received power \eqref{eq:lin}, both evaluated at the ray-traced $P_{\LoS}=0.495$. Arrows show $\pm\sigma$ around each mode. (c) Outage probability against altitude for a 118~dB link budget, from the ray trace (markers), the mixture \eqref{eq:mix}, and the decibel average \eqref{eq:dbavg} with the exact $P_{\LoS}$ or with the published sigmoid \eqref{eq:plos}. Triangles mark the optima at 1764~m and 778~m.}
\label{fig:L3}
\vspace{-0.8\baselineskip}
\end{figure*}

Instead, this letter examines one step that follows the construction. In particular, having obtained $P_{\LoS}$, a large body of work \cite{alhourani2014wcl,mozaffari2015,boryaliniz2016,alzenad2017,alzenad2018,wang2024,vavoulas2026} forms a single mean path loss as
\begin{equation}
\overline{\PL}=\PL_{\mathrm{FS}}+P_{\LoS}\,\eta_{\LoS}+\left(1-P_{\LoS}\right)\eta_{\NLoS},
\label{eq:dbavg}
\end{equation}
in which decibel values are weighted by probabilities. The value returned by \eqref{eq:dbavg} is then treated as the path loss of the link and used to define a coverage radius, an optimal altitude, or an outage probability. However, as illustrated in Fig.~\ref{fig:concept}(a), a link is either in the LoS or in the NLoS state, and never in between. The conditional path-loss distribution is therefore bimodal, with modes at $\PL_{\mathrm{FS}}+\eta_{\LoS}$ and $\PL_{\mathrm{FS}}+\eta_{\NLoS}$, and the weighted value falls in the valley between the modes, where no link lies (Fig.~\ref{fig:L3}(b)). We derive a closed-form expression for the resulting error and show that a two-state mixture with the same inputs removes it.

The weighted form \eqref{eq:dbavg} was introduced in~\cite{alhourani2014wcl} as a deliberate simplification, whose authors state that they use the mean of the excess loss rather than its random behavior. It has since been adopted, usually without this caveat, in drone small-cell deployment~\cite{mozaffari2015}, three-dimensional aerial base station placement~\cite{boryaliniz2016,alzenad2017,alzenad2018}, reinforcement-learning placement~\cite{wang2024}, and coverage of convex regions~\cite{vavoulas2026}. Table~\ref{tab:instances} lists these works with the quantity that each derives from \eqref{eq:dbavg}. In each case, \eqref{eq:dbavg} is a reasonable first-order simplification, and the aim of this letter is to quantify its cost.

Recent work has modeled the excess gain as a two-component Gaussian mixture and computed coverage from the mixture distribution~\cite{alhomssi2022}. In addition, our earlier survey cautioned that \eqref{eq:dbavg} should be used carefully in system-level analysis of throughput and outage, and that path-loss variability should be added to it~\cite{khawaja2019}. That caution was qualitative, and this letter quantifies it. Later refinements have improved the geometric fidelity of the LoS model~\cite{alhourani2020,saboor2024,chen2024}, but none of them addresses the averaging step that follows it, so the results of this letter apply to any LoS probability model.

\begin{table}[!t]
\centering
\footnotesize
\setlength{\tabcolsep}{3.5pt}
\caption{Inventory of the decibel-domain average \eqref{eq:dbavg} in the aerial coverage literature and the quantity that it drives.}
\label{tab:instances}
\begin{tabular}{@{}lll>{\raggedright\arraybackslash}p{0.95in}@{}}
\toprule
Work & Venue, year & Eq. & Averaged value drives\\
\midrule
Al-Hourani \emph{et al.}~\cite{alhourani2014wcl} & WCL, 2014   & (9) & coverage radius, optimal altitude\\
Mozaffari \emph{et al.}~\cite{mozaffari2015}     & GLOBECOM, 2015 & (4) & coverage radius, optimal altitude\\
Bor-Yaliniz \emph{et al.}~\cite{boryaliniz2016}  & ICC, 2016   & (2) & per-user path-loss constraint\\
Alzenad \emph{et al.}~\cite{alzenad2017}         & WCL, 2017   & (3) & coverage radius, required power\\
Alzenad \emph{et al.}~\cite{alzenad2018}         & WCL, 2018   & (3) & per-QoS-class coverage disc\\
Wang and Farooq~\cite{wang2024}                  & IoT-J, 2024 & (2) & SNR, rate, learning reward\\
Vavoulas \emph{et al.}~\cite{vavoulas2026}       & WCL, 2026   & (11)& placement objective, altitude\\
\bottomrule
\end{tabular}
\vspace{-3mm}
\end{table}

\section{Averaging Error and Mixture Model}

Let $\xi\in\{\LoS,\NLoS\}$ denote the link state and $g$ the linear channel gain. The expected value of $g$ corresponds physically to the mean received power, and it is obtained by averaging the two states in the linear domain,
\begin{equation}
\mathbb{E}[g]=P_{\LoS}\,10^{-\frac{\PL_{\mathrm{FS}}+\eta_{\LoS}}{10}}+\left(1-P_{\LoS}\right)10^{-\frac{\PL_{\mathrm{FS}}+\eta_{\NLoS}}{10}} .
\label{eq:lin}
\end{equation}
Converting \eqref{eq:lin} back to decibels gives a value that differs from \eqref{eq:dbavg} because the logarithm is concave. The two coincide only when $\eta_{\LoS}=\eta_{\NLoS}$, in which case no averaging is needed.

This difference has a closed form, which predicts the cost of the approximation in advance. Subtracting the decibel equivalent of \eqref{eq:lin} from \eqref{eq:dbavg} and writing $p$ for $P_{\LoS}$ gives the averaging error
\begin{equation}
\begin{split}
\Delta={}&p\,\eta_{\LoS}+(1{-}p)\,\eta_{\NLoS}\\
&+10\log_{10}\!\left[p\,10^{-\eta_{\LoS}/10}+(1{-}p)\,10^{-\eta_{\NLoS}/10}\right],
\end{split}
\label{eq:gap}
\end{equation}
which depends only on $p$ and on the two excess losses, since $\PL_{\mathrm{FS}}$ cancels. The error $\Delta$ becomes zero at $p=0$ and at $p=1$, where one state is certain and there is no averaging to do. Fig.~\ref{fig:concept}(b) shows $\Delta$ between these limits. For the urban mode separation of 19~dB, the error peaks at 8.4~dB when $P_{\LoS}=0.22$. Reducing the separation to 10~dB caps the error at 2.7~dB, and raising it to 30~dB pushes the peak to 17.3~dB. The maximum value of \eqref{eq:gap} therefore does not occur at $p=0.5$, because the linear average is pulled towards the stronger state, and the maximum moves towards a lower value of $p$ as the modes separate.

Two limiting cases of \eqref{eq:gap} clarify its behavior. Writing $\Deta=\eta_{\NLoS}-\eta_{\LoS}$, a second-order expansion for small $\Deta$ and the dominant term for large $\Deta$ give
\begin{equation}
\Delta\approx
\begin{cases}
\dfrac{\ln 10}{20}\,p(1-p)\,\Deta^{2}, & \Deta\ \text{small},\\[6pt]
(1-p)\,\Deta+10\log_{10}p, & \Deta\ \text{large}.
\end{cases}
\label{eq:limits}
\end{equation}
The first line is the second-order (delta-method) approximation~\cite{rice2007} of the Jensen gap of a two-point distribution with variance $p(1-p)\Deta^{2}$. For nearby modes (small $\Deta$), the error therefore grows as the square of their separation and is symmetric in $p$. It is accurate to within 5\% at $\Deta=3$~dB and $p=0.5$, but it overestimates the exact value by 59\% at $\Deta=19$~dB, the difference between the urban excess losses of~\cite{alhourani2014wcl} that Section~\ref{sec:rt} uses. In that regime, the second line of \eqref{eq:limits} applies and the error becomes linear in $\Deta$. The Appendix derives both lines. Setting its derivative to zero places the peak at $p\approx10/(\Deta\ln 10)$, that is, 0.23 at 19~dB and 0.14 at 30~dB, against exact peaks at 0.22 and 0.14. At $\Deta=30$~dB, the large-separation form reproduces the exact peak of 17.3~dB to within 0.1~dB.

Regardless of whether \eqref{eq:dbavg} or \eqref{eq:lin} is used, a single path-loss value, e.g., as used in~\cite{alhourani2014wcl,mozaffari2015,boryaliniz2016,alzenad2017,alzenad2018,wang2024,vavoulas2026}, cannot describe a bimodal link in an outage calculation. A coverage study requires the probability that the path loss exceeds a budget $\PL_{\max}$, which follows from the law of total probability over the two states,
\begin{equation}
P_{\mathrm{out}}=\sum_{\xi\in\{\LoS,\NLoS\}}P_{\xi}\;Q\!\left(\frac{\PL_{\max}-\PL_{\mathrm{FS}}-\eta_{\xi}}{\sigma_{\xi}}\right),
\label{eq:mix}
\end{equation}
where $\sigma_{\xi}$ is the shadowing standard deviation in state $\xi$ and $Q(\cdot)$ is the Gaussian tail function. Expression~\eqref{eq:mix} keeps the two modes separate and weights their individual tail probabilities. A link budget typically lies between the two modes, so the NLoS mode contributes almost all of the outage, and collapsing the modes first removes this structure.

The same argument applies to any nonlinear function of the link state, and the rate is the most common such function in coverage studies. With $\gamma_{\xi}$ the signal-to-noise ratio (SNR) in state $\xi$ and $X$ the shadowing in decibels, the ergodic spectral efficiency is
\begin{equation}
\mathbb{E}[C]=\sum_{\xi}P_{\xi}\,
\mathbb{E}_{X}\!\left[\log_{2}\!\left(1+\gamma_{\xi}10^{-X/10}\right)\right],
\label{eq:rate}
\end{equation}
which again mixes after the nonlinearity rather than before it. Collapsing the states first replaces \eqref{eq:rate} by a single logarithm evaluated at the averaged path loss. Fig.~\ref{fig:concept}(c) shows that, at an LoS SNR of 20~dB, the use of \eqref{eq:dbavg} understates the ergodic rate by 10.3\% at $P_{\LoS}=0.495$ and by up to 12.5\% near $P_{\LoS}=0.29$. The sign of this error is opposite to that of the outage error. Decibel averaging makes a link look weaker in mean power, so the rate looks worse, whereas the same step makes the outage look better by removing the NLoS tail.

\section{Ray-Traced Evaluation}
\label{sec:rt}

To separate the averaging error from any error in $P_{\LoS}$ itself, we build an explicit city rather than relying on a fitted model. Buildings are placed on the ITU-R P.1410 grid with built-up fraction $\alpha=0.3$, density $\beta=500$~buildings/km$^{2}$, and Rayleigh height parameter $\gamma=15$~m, giving a 44.7~m pitch and 24.5~m footprints~\cite{itu1410}. Ground users at a height of 1.5~m are placed uniformly in the streets, and every link is ray tested against the building heights to obtain its exact blockage state. This geometric ray tracing determines only whether the direct ray is blocked. Reflection and diffraction enter through the excess losses and the shadowing below. Link states are converted to path loss at a 3.5~GHz carrier, with the urban excess losses $\eta_{\LoS}=1$~dB and $\eta_{\NLoS}=20$~dB of~\cite{alhourani2014wcl,boryaliniz2016}, and shadowing standard deviations $\sigma_{\LoS}=4$~dB and $\sigma_{\NLoS}=8$~dB. An outage probability from \eqref{eq:dbavg} also needs a shadowing spread, which \eqref{eq:dbavg} does not define because it represents both states by a single mean. We therefore model the path loss of the averaged model as one Gaussian with mean $\overline{\PL}$ and standard deviation $\bar{\sigma}=P_{\LoS}\sigma_{\LoS}+(1-P_{\LoS})\sigma_{\NLoS}$, which interpolates the two spreads in the same way as \eqref{eq:dbavg} interpolates the two means. Its outage is then $Q\big((\PL_{\max}-\overline{\PL})/\bar{\sigma}\big)$. With this choice, the averaged model and the mixture \eqref{eq:mix} use the same inputs and differ only in how they combine the two states.

Fig.~\ref{fig:L3}(a) compares three LoS probability curves. The markers are the ray-traced $P_{\LoS}$ at horizontal ranges of 200~m, 600~m, and 1500~m. They agree within 0.06 at any elevation angle, so the elevation-only form \eqref{eq:plos} suits this city. The dashed curve is \eqref{eq:plos} with the urban pair $(a,b)=(9.61,0.16)$ reported in~\cite{alhourani2014wcl,boryaliniz2016} and used by the studies in Table~\ref{tab:instances}. It overestimates the traced $P_{\LoS}$ by up to 0.33, at $\theta=40^{\circ}$. The solid curve is \eqref{eq:plos} refitted to the trace, with $(a,b)=(5.70,0.063)$ and a root-mean-square error of 0.043.

The published pair closely follows the ITU-R P.1410 closed form~\cite{itu1410}, which gives $P_{\LoS}=0.79$ against the traced 0.495 for the link of Section~\ref{sec:link}. The difference comes from the position of the user. The closed form places the first building half a spacing from the user, whereas a street-level user in the explicit grid always has a wall within one street width of 20.2~m. A wall at distance $x$ blocks the ray when it is taller than $x\tan\theta$, which for Rayleigh heights has probability $\exp(-x^{2}\tan^{2}\theta/2\gamma^{2})$, or 0.51 at $x=10$~m and $\theta=60^{\circ}$.

The key control in this experiment is that the averaged model \eqref{eq:dbavg} is supplied with the ray-traced $P_{\LoS}$ rather than with a fitted sigmoid. Any discrepancy that remains is therefore attributable to the averaging step alone, and not to a parameter error. For reference, we also evaluate \eqref{eq:dbavg} with the published sigmoid of Fig.~\ref{fig:L3}(a).

\section{Numerical Results}

\subsection{Path Loss Distribution for a Representative Link}
\label{sec:link}

In Fig.~\ref{fig:L3}(b), we show the path-loss distribution for a representative link at 500~m range and 300~m altitude, where the elevation angle is 30.8$^{\circ}$ and $\PL_{\mathrm{FS}}=98.6$~dB. Ray tracing gives $P_{\LoS}=0.495$ at this geometry, and the distribution is strongly bimodal, with modes at $\PL_{\mathrm{FS}}+\eta_{\LoS}=99.6$~dB and $\PL_{\mathrm{FS}}+\eta_{\NLoS}=118.6$~dB, whose widths are set by $\sigma_{\LoS}$ and $\sigma_{\NLoS}$. The decibel-weighted mean path loss of \eqref{eq:dbavg}, supplied with that exact probability, is 109.2~dB and lies in the valley between the modes.

This placement has three consequences. First, only 6.2\% of the 60\,000 ray-traced links have a path loss within 2~dB of 109.2~dB, and only 19.1\% within 5~dB. The averaged value thus describes almost none of the links it summarizes. Second, the mean received power from \eqref{eq:lin} corresponds to a path loss of 102.6~dB, so the averaged path loss is 6.6~dB too high. Equation~\eqref{eq:gap} returns the same 6.6~dB from $P_{\LoS}$ and the two excess losses alone, as the star in Fig.~\ref{fig:concept}(b) shows. Third, a link budget set at 109.2~dB would be met by 99\% of the LoS links but by only 12\% of the NLoS links, so the averaged value predicts the coverage of neither state.

Log-normal shadowing introduces a separate effect that is sometimes confused with the averaging error. For shadowing $X\sim\mathcal{N}(0,\sigma^{2})$ in decibels, $\mathbb{E}[10^{-X/10}]=\exp(\sigma^{2}\ln^{2}10/200)$, so shadowing raises the mean received power by a further $\sigma^{2}\ln 10/20$ decibels. This lift is 1.8~dB at $\sigma=4$~dB and 7.4~dB at $\sigma=8$~dB. It is a property of the shadowing distribution rather than of the two-state average, and \eqref{eq:lin} excludes it so that the averaging error is isolated. When the lift is included analytically, the gap at this geometry increases to 8.6~dB.

\subsection{Coverage and the Optimal Altitude}

Fig.~\ref{fig:L3}(c) shows the effect of the averaging in \eqref{eq:dbavg} on a coverage calculation. The outage probability for a 118~dB budget is evaluated as the aerial platform climbs from 40~m to 4000~m. This range is wide enough that the ray-traced outage rises again beyond its minimum of 0.096 at 1764~m. Supplied with the exact ray-traced $P_{\LoS}$, the mixture of \eqref{eq:mix} reproduces the ray-traced outage to within 0.004 and finds the same optimum altitude. The decibel average, given the same exact $P_{\LoS}$, places the optimum at 778~m, a factor of 2.3 too low, and understates the outage by up to 0.198.

These two curves receive the same $P_{\LoS}$ and differ only in how they combine the two states. Their difference therefore cannot come from a mis-fitted sigmoid, and it remains after any improvement to the LoS probability model. The third curve in Fig.~\ref{fig:L3}(c) shows that the published sigmoid adds a parameter error to the averaging error. With this sigmoid, the decibel average places the optimum at 517~m, and its largest outage error rises to 0.269.

The altitude error has a simple physical explanation. Climbing raises $P_{\LoS}$, which moves \eqref{eq:dbavg} smoothly towards the LoS mode, so the averaged path loss falls below the 118~dB budget already at a modest altitude. The ray-traced outage, by contrast, is dominated by the NLoS links, whose share falls only as fast as $1-P_{\LoS}$ does. At the 778~m chosen by the decibel average, the ray-traced outage is still 0.161 (Table~\ref{tab:env}), against the minimum of 0.096 at 1764~m.

\begin{table}[!t]
\centering
\footnotesize
\setlength{\tabcolsep}{2.6pt}
\caption{Controlled comparison in the four ITU-R P.1410 environments, at 500~m range and 300~m altitude for $P_{\LoS}$ and $\Delta$ from \eqref{eq:gap}, and over 40--4000~m altitude for the outage columns. $h^{\star}$ is the ray-traced optimum altitude and $h^{\mathrm{dB}}$ the altitude chosen by the decibel average. $P_{\mathrm{out}}^{\star}$ is the ray-traced minimum outage, $P_{\mathrm{out}}^{\mathrm{dB}}$ the ray-traced outage at $h^{\mathrm{dB}}$, and $\epsilon_{\max}$ the largest outage error of the decibel average over altitude.}
\label{tab:env}
\begin{tabular}{@{}lrrrrrrr@{}}
\toprule
Environment & $P_{\LoS}$ & $\Delta$ (dB) & $h^{\star}$ (m) & $h^{\mathrm{dB}}$ (m) & $P_{\mathrm{out}}^{\star}$ & $P_{\mathrm{out}}^{\mathrm{dB}}$ & $\epsilon_{\max}$\\
\midrule
Suburban        & 0.84 & 2.3 & 1438 & 421  & 0.028 & 0.065 & 0.19\\
Urban           & 0.49 & 6.6 & 1764 & 778  & 0.097 & 0.161 & 0.20\\
Dense urban     & 0.29 & 8.3 & 2165 & 1171 & 0.171 & 0.232 & 0.18\\
High-rise urban & 0.08 & 7.1 & 2656 & 466  & 0.378 & 0.506 & 0.23\\
\bottomrule
\end{tabular}
\end{table}

\subsection{Results in Four ITU-R P.1410 Environments}

The results above are obtained for a single city, and their magnitudes may depend on the urban parameters. We therefore repeated the controlled comparison in the four standard ITU-R P.1410 environments~\cite{itu1410}, from suburban to high-rise, with the same link budget, excess losses, and geometry, and in an independent ray-traced realization. Table~\ref{tab:env} summarizes the results, and its urban row reproduces Fig.~\ref{fig:L3} to within the Monte Carlo error.

Three regularities follow from Table~\ref{tab:env}. First, in every environment, the mixture \eqref{eq:mix} supplied with the exact $P_{\LoS}$ selects the same optimum altitude $h^{\star}$ as the ray trace. Its outage differs from the ray-traced outage by at most 0.0055 at any altitude. Second, the decibel average selects an altitude $h^{\mathrm{dB}}$ that is lower than $h^{\star}$ by a factor between 1.8 and 5.7. Its largest outage error $\epsilon_{\max}$ stays between 0.18 and 0.23, regardless of how the city is built. Third, the averaging error $\Delta$ at the representative geometry follows \eqref{eq:gap} through $P_{\LoS}$ alone. It is smallest in the suburban city, where $P_{\LoS}=0.84$ puts the link almost entirely in one state. It is largest in the dense urban city, where $P_{\LoS}=0.29$ lies near the peak of the 19~dB curve in Fig.~\ref{fig:concept}(b).

For a designer, the operational cost is the more useful measure. An aerial base station placed at $h^{\mathrm{dB}}$ has a ray-traced outage $P_{\mathrm{out}}^{\mathrm{dB}}$ of 0.065 instead of $P_{\mathrm{out}}^{\star}=0.028$ in the suburban city, a factor of 2.3. In the high-rise city, the outage is 0.506 instead of 0.378. The penalty is therefore not confined to one urban parameter set, and it does not shrink as the environment becomes more open.

\section{Conclusion}

Weighting the LoS and NLoS path losses by their probabilities in the decibel domain produces a value that describes no physical link. Even when it is given the exact LoS probability of a ray-traced ITU-R P.1410 city, this average lies 6.6~dB from the path loss of the mean received power. It also places the coverage-optimal altitude 1.8 to 5.7 times too low in four standard environments. Three simple changes remove the error. A mean received power should be computed by averaging the linear gain as in \eqref{eq:lin}. An outage or coverage probability should be computed by mixing the conditional tails as in \eqref{eq:mix}, and a rate by mixing after the logarithm as in \eqref{eq:rate}. The excess-loss and shadowing values used here are representative rather than site-specific, although \eqref{eq:gap} and \eqref{eq:limits} show how the error scales with the mode separation. A measured dataset that records the link state together with the path loss is the natural next test.

\section*{Appendix: Derivation of \eqref{eq:limits}}
Substituting $\eta_{\NLoS}=\eta_{\LoS}+\Deta$ gives $10^{-\eta_{\NLoS}/10}=10^{-\eta_{\LoS}/10}\,10^{-\Deta/10}$, so both terms of the bracket in \eqref{eq:gap} share the factor $10^{-\eta_{\LoS}/10}$. With $10\log_{10}(xy)=10\log_{10}x+10\log_{10}y$ and $10\log_{10}10^{-\eta_{\LoS}/10}=-\eta_{\LoS}$, the logarithm in \eqref{eq:gap} becomes $-\eta_{\LoS}+10\log_{10}[p+(1-p)\,10^{-\Deta/10}]$, while the first line of \eqref{eq:gap} becomes $\eta_{\LoS}+(1-p)\Deta$. Adding the two, $\eta_{\LoS}$ cancels and
\begin{equation}
\Delta=(1-p)\,\Deta+10\log_{10}\!\left[p+(1-p)\,10^{-\Deta/10}\right].
\label{eq:app}
\end{equation}
For small $\Deta$, let $\beta=\ln 10/10$, so that $10^{-\Deta/10}=e^{-\beta\Deta}$ and $10\log_{10}(\cdot)=\ln(\cdot)/\beta$. The bracket in \eqref{eq:app} is then $\mathbb{E}[e^{tB}]$ for $B\sim\mathrm{Bernoulli}(1-p)$ at $t=-\beta\Deta$, whose logarithm expands as $t(1-p)+t^{2}p(1-p)/2+O(t^{3})$. The logarithmic term of \eqref{eq:app} is therefore $-(1-p)\Deta+(\beta/2)\,p(1-p)\,\Deta^{2}+O(\Deta^{3})$. Its linear part cancels the first term, which leaves the first line of \eqref{eq:limits}. For large $\Deta$, the bracket tends to $p$, and \eqref{eq:app} tends to the second line, with the exact remainder $10\log_{10}\!\big[1+\tfrac{1-p}{p}\,10^{-\Deta/10}\big]$.

\newpage
\section*{Acknowledgment}
The author used Anthropic's Claude Code and Claude Cowork to assist with simulation code development, data analysis, and drafting and editing of the manuscript text. All AI-assisted content was independently verified by the author; content that could not be verified was revised or discarded. The author takes full responsibility for the work.

\bibliographystyle{IEEEtran}
\bibliography{refs}

@article{khawaja2019,author={W. Khawaja and I. Guvenc and D. W. Matolak and U.-C. Fiebig and N. Schneckenburger},title={A Survey of Air-to-Ground Propagation Channel Modeling for Unmanned Aerial Vehicles},journal={IEEE Commun. Surveys Tuts.},volume={21},number={3},pages={2361--2391},month={3rd Quart.},year={2019}}

@article{alhourani2014wcl,author={A. Al-Hourani and S. Kandeepan and S. Lardner},title={Optimal {{LAP}} Altitude for Maximum Coverage},journal={IEEE Wireless Commun. Lett.},volume={3},number={6},pages={569--572},month=dec,year={2014}}

@techreport{itu1410,author={{ITU-R}},title={Propagation Data and Prediction Methods Required for the Design of Terrestrial Broadband Radio Access Systems Operating in a Frequency Range from 3 to 60 {{GHz}}},institution={International Telecommunication Union},number={Recommendation P.1410-6},month=aug,year={2023}}

@inproceedings{mozaffari2015,author={M. Mozaffari and W. Saad and M. Bennis and M. Debbah},title={Drone Small Cells in the Clouds: Design, Deployment and Performance Analysis},booktitle={Proc. IEEE Global Commun. Conf. (GLOBECOM)},address={San Diego, CA},pages={1--6},month=dec,year={2015}}

@inproceedings{boryaliniz2016,author={R. I. Bor-Yaliniz and A. El-Keyi and H. Yanikomeroglu},title={Efficient {3-D} Placement of an Aerial Base Station in Next Generation Cellular Networks},booktitle={Proc. IEEE Int. Conf. Commun. (ICC)},address={Kuala Lumpur, Malaysia},pages={1--5},month=may,year={2016}}

@article{alzenad2017,author={M. Alzenad and A. El-Keyi and F. Lagum and H. Yanikomeroglu},title={{3-D} Placement of an Unmanned Aerial Vehicle Base Station ({{UAV-BS}}) for Energy-Efficient Maximal Coverage},journal={IEEE Wireless Commun. Lett.},volume={6},number={4},pages={434--437},month=aug,year={2017}}

@article{alzenad2018,author={M. Alzenad and A. El-Keyi and H. Yanikomeroglu},title={{3-D} Placement of an Unmanned Aerial Vehicle Base Station for Maximum Coverage of Users with Different {{QoS}} Requirements},journal={IEEE Wireless Commun. Lett.},volume={7},number={1},pages={38--41},month=feb,year={2018}}

@article{wang2024,author={Y. Wang and J. Farooq},title={Deep-Reinforcement-Learning-Based Placement for Integrated Access Backhauling in {{UAV}}-Assisted Wireless Networks},journal={IEEE Internet Things J.},volume={11},number={8},pages={14727--14738},month=apr,year={2024}}

@article{vavoulas2026,author={A. Vavoulas and K. K. Delibasis and H. G. Sandalidis and G. Nousias and N. Vaiopoulos},title={Efficient {{UAV}} Coverage in Large Convex Quadrilateral Areas with Elliptical Footprints},journal={IEEE Wireless Commun. Lett.},volume={15},pages={1479--1483},year={2026}}

@article{alhomssi2022,author={B. Al Homssi and A. Al-Hourani},title={Optimal Beamwidth and Altitude for Maximal Uplink Coverage in Satellite Networks},journal={IEEE Wireless Commun. Lett.},volume={11},number={4},pages={771--775},month=apr,year={2022}}

@article{alhourani2020,author={A. Al-Hourani},title={On the Probability of Line-of-Sight in Urban Environments},journal={IEEE Wireless Commun. Lett.},volume={9},number={8},pages={1178--1181},month=aug,year={2020}}

@article{saboor2024,author={A. Saboor and E. Vinogradov and Z. Cui and A. Al-Hourani and S. Pollin},title={A Geometry-Based Modelling Approach for the Line-of-Sight Probability in {{UAV}} Communications},journal={IEEE Open J. Commun. Soc.},volume={5},pages={364--378},year={2024}}

@article{chen2024,author={L. Chen and W. Zhang and M. A. Kishk and M.-S. Alouini},title={Correlation of Line-of-Sight Probabilities in Aerial-Terrestrial Communications: Modeling, Analysis, and Application},journal={IEEE Trans. Veh. Technol.},volume={73},number={5},pages={7049--7065},month=may,year={2024}}

@article{zeng2019,author={Y. Zeng and Q. Wu and R. Zhang},title={Accessing from the Sky: A Tutorial on {{UAV}} Communications for 5{{G}} and Beyond},journal={Proc. IEEE},volume={107},number={12},pages={2327--2375},month=dec,year={2019}}

@book{rice2007,author={J. A. Rice},title={Mathematical Statistics and Data Analysis},edition={3rd},publisher={Brooks/Cole},address={Belmont, CA},year={2007},note={Sec.~4.6}}

\end{document}